\documentclass[conference]{IEEEtran}

\usepackage{graphicx}
\usepackage{amsmath,amssymb}
\usepackage{siunitx}
\usepackage{upgreek}
\usepackage[noend]{algpseudocode}
\usepackage{algorithm}
\usepackage{multirow}
\usepackage{url}
\usepackage{balance}

\usepackage{xcolor}
\usepackage{soul}

\ifCLASSINFOpdf
\else
\fi

\begin{document}
%
\title{Hyperspectral Band Selection for Multispectral Image Classification with Convolutional Networks}



\author{\IEEEauthorblockN{Giorgio Morales and John Sheppard}
\IEEEauthorblockA{Gianforte School of Computing \\
Montana State University\\
Bozeman, MT 59717}
\and
\IEEEauthorblockN{Riley Logan and Joseph Shaw}
\IEEEauthorblockA{Department of Electrical and Computer Engineering \\
Montana State University
\\
Bozeman, MT 59717}}
        
        

\maketitle

\begin{abstract}

In recent years, Hyperspectral Imaging (HSI) has become a powerful source for reliable data in applications such as remote sensing, agriculture, and biomedicine. However, hyperspectral images are highly data-dense and often benefit from methods to reduce the number of spectral bands while retaining the most useful information for a specific application. We propose a novel band selection method to select a reduced set of wavelengths, obtained from an HSI system in the context of image classification. Our approach consists of two main steps: the first utilizes a filter-based approach to find relevant spectral bands based on a collinearity analysis between a band and its neighbors. This analysis helps to remove redundant bands and dramatically reduces the search space. The second step applies a wrapper-based approach to select bands from the reduced set based on their information entropy values, and trains a compact Convolutional Neural Network (CNN) to evaluate the performance of the current selection. 
We present classification results obtained from our method and compare them to other feature selection methods on two hyperspectral image datasets. Additionally, we use the original hyperspectral data cube to simulate the process of using actual filters in a multispectral imager. We show that our method produces more suitable results for a multispectral sensor design.\footnote{This paper is a preprint (accepted to appear in the International Joint Conference on Neural Networks 2021). IEEE copyright notice. 2021 IEEE. Personal use of this material is permitted. Permission from IEEE must be obtained for all other uses, in any current or future media, including reprinting/republishing this material for advertising or promotional purposes, creating new collective works, for resale or redistribution to servers or lists, or reuse of any copyrighted.
}
\end{abstract}

\section{Introduction}

Optical remote sensing systems have a long history of collecting image data for many diverse applications, ranging from lab-based analysis of food quality and safety \cite{Wu2013e} to space-based contributions to archaeology \cite{Giardino2011}. The cornerstone of these systems is the exploitation of spatial and spectral information contained within the captured imagery. Though spatial information within an image can provide useful information, spectral data plays a central role in identifying and classifying objects in a scene. To address the need for rich spectral information, optical remote sensing systems come in many forms, ranging from simple multispectral imaging (MSI) systems  \cite{MSI1,MSI2} to hyperspectral imagering (HSI) systems \cite{Wu2013e,HSI2}. In an MSI system, several distinct spectral bands are captured, often outside the visible spectrum. These systems are useful for capturing information in known areas of the spectrum. For example, if an application requires the detection of vegetation, a multispectral imager may only be required to capture the reflectance at \SI{680}{\nano\meter} (red) and \SI{800}{\nano\meter} (near-infrared), two commonly used spectral channels, to capture the chlorophyll content \cite{Xue2017}. In contrast to MSI,  HSI systems often capture hundreds of contiguous spectral bands.

Though powerful, the spectrally dense images captured during HSI come at the price of high data density, large file size, and increased computational complexity, which represent computational limitations when storing and processing these types of images. Returning to the example above, instead of capturing two distinct spectral bands to detect vegetation, a hyperspectral imager would capture data from hundreds of bands surrounding the wavelengths of interest. In such situations, the complexity introduced by the HSI system may be unnecessary if similar detection results can be achieved with fewer spectral channels. However, in many applications, relevant wavelengths are not known \textit{a priori}. The ability to determine the most important wavelengths in a hyperspectral image would greatly simplify the data capture and processing requirements.
Namely, it would enable using multispectral imagers in place of hyperspectral imagers, greatly reducing complexity and cost. Unfortunately, selecting salient wavelengths from an HSI system is not a trivial task.

In this paper, we propose a feature selection method to identify the most relevant spectral bands given an HSI classification problem.
Our feature selection method consists of two steps: the first is a novel pre-selection method that we call inter-band redundancy analysis (IBRA). It assesses the degree of collinearity between each spectral band and its neighbors in order to approximate the minimum number of bands we need to move away from a band to find spectral bands with sufficiently distinct information. The distribution of this distance metric across the spectrum helps us to identify a reduced set of independent bands that act as the centroid of their corresponding regions in the spectrum. The second step is called greedy spectral selection (GSS) and consists of calculating the information entropy of each pre-selected band to rank its relevance. Then, we train a classifier using the top $k$ pre-selected bands (where $k$ is the number of desired bands) sorted according to their corresponding entropy. Finally, we remove from our selection the band that shows the most severe indication of multicollinearity and repeat the process taking into account the next available band to verify if the classification performance improves.

Having selected a reduced number of spectral bands from the original hyperspectral image, we train a new classification model based on a reduced-parameter convolutional neural network \cite{hyper3dnet} that achieves high performance. 
We hypothesize that it is possible to apply the combination of inter-band redundancy analysis and greedy spectral selection to select a small number of wavelengths ($\sim$5--10) that will lead us to train more efficient HSI classifiers than the compared methods.


\section{Related Work} \label{background}

Several dimensionality reduction techniques have been applied in the past as a natural pre-processing step for HSI classification problems. This is done to avoid unnecessarily high time complexity when processing large volumes of data and to reduce redundancy of the data, which could impair the performance of a classifier \cite{review}. Dimensionality reduction techniques rely on feature extraction or feature selection approaches; the former apply linear or non-linear transformations to extract specific features from the original data, while the latter select the most useful subset of the features of the data without transforming them.

Among feature extraction methods, principal component analysis (PCA) and its variants (e.g. folded-PCA and kernel PCA) are some of the most commonly used methods to remove spectral redundancy and reduce the dimensionality of the raw data \cite{Uddin}. On the other hand, feature selection methods select a subset of spectral bands without modifying the data or projecting it into a new basis. The aim of this work is to identify which spectral wavelengths from the original hyperspectral spectrum are most responsive or relevant for a particular classification task without modifying the data, which is why we prefer feature selection methods over feature extraction methods. Additionally, identifying a reduced subset of relevant spectral bands allows for a better understanding of the optical properties of the materials and provides information that is useful when designing cheaper task-specific multispectral imagers. For example, if an application demands the identification of a certain type of plant, a feature selection method will identify the wavelengths that change most due to absorption from the unique pigmentation in the plant.

Given the advantages of feature selection, several methods have been proposed for hyperspectral image classification, one of the most common being ranking-based methods. These methods estimate the importance of each spectral band using such metrics as the variance inflation factor (VIF) \cite{castaldi} in order to select the top-ranked bands. We will use the idea of calculating the VIF value to measure collinearity, but the spectral bands will not be ranked based on this simple measure alone.  
Other methods use the estimated band relevance as part of an optimization approach, as proposed by Wang et al. \cite{cluster}, where the optimal clustering framework (OCF) separates the data into clusters, ranks them according to a selected measure (e.g. information entropy), and selects those with higher rank values. Furthermore, Wang et al. \cite{fast} also proposed a fast neighborhood grouping method for hyperspectral band selection (FNGBS) that partitions the data into several groups using Euclidean distance as a similarity measure, and then obtains the most relevant and informative bands using local density and information entropy measures.

In recent years, two feature selection approaches for HSI classification have gained more attention: partial least squares discriminant analysis (PLS-DA) \cite{Feng2019} and genetic algorithms \cite{Shaoping}. For instance, a recent method for bandwidth selection is known as Histogram Assisted Genetic Algorithm for Reduction in Dimensionality (HAGRID) \cite{Neil}. This method maintains a population of index vectors identifying some specific number of wavelengths and fits a Gaussian mixture model to the converged population to identify the main wavelengths with their associated filter bandwidths.

Alternatively, some model-based approaches have been proposed in the context of deep learning. 
For instance, Taherkhani \textit{et al.} \cite{Fariborz} proposed regularizing the convolutional filters of the first layer of the convolutional neural network (CNN) using a group LASSO algorithm in order to sparsify the redundant spectral bands. 
Similar attempts, although not explicit feature selection methods, have been carried out in works such as \cite{Fang_2019}, and \cite{Gao2019}, where a spectral-wise attention mechanism in the form of a fully-connected layer is applied to the inner convolutional layers of the network with the objective of emphasizing informative spectral features and suppressing less useful spectral features. 


\section{Materials and Methods} \label{design}

\subsection{Datasets}

We will use two datasets: an in-greenhouse controlled hyperspectral image dataset called ``Kochia'' and a well-known remote sensing HSI dataset called ``Indian Pines'' (IP). 

The Kochia dataset consists of images of the weed kochia (\textit{Bassia scoparia}) that were collected and analyzed by Scherrer \textit{et al.} \cite{Bryan} with the aim of learning to differentiate between three different classes of herbicide-resistance of this weed: 1) herbicide-susceptible, 2) glyphosate-resistant, and 3) dicamba-resistant, where glyphosate and dicamba are two components commonly found in commercial herbicides. The images were captured using a Resonon Pika L hyperspectral imager with 300 spectral channels across a spectral range of 387.12 nm to 1,023.50 nm, resulting in a spectral resolution of approximately 2.12 nm. The kochia samples were illuminated using diffuse sunlight in a greenhouse setting. A total of 76 hyperspectral images of kochia with varying ages and spatial resolutions were captured at the Montana State University Southern Agricultural Research Center (SARC). Each image contains three kochia leaves of the same herbicide-resistance class with a height of 900 pixels and width ranging from 700--1,200 pixels.

The Indian Pines dataset \cite{IndianPines} is an aerial $145 \times 145$ - pixel image of the Indian Pines site in Northwestern Indiana. It was acquired using the Airborne Visible / Infrared Imaging Spectrometer (AVIRIS) sensor \cite{aviris} and it originally had 224 spectral bands in the wavelength range 380-2,510 nm, resulting in a spectral resolution of approximately 9.5 nm. The number of bands was reduced to 200 after removing 24 bands covering the region of water absorption. The data are divided into 16 classes containing agriculture, forest, and other natural perennial vegetation.

\subsection{Data Pre-Processing}

Images of kochia leaves were captured in raw digital numbers recorded by the Pika L hyperspectral imager, meaning the data cubes required pre-processing before they could be analyzed. For these experiments, data pre-processing was limited to reflectance correction for the Kochia dataset. To accomplish this, we converted the raw digital numbers to reflectance values using a 99\% Spectralon panel as a reflectance reference. Specifically, the calculation of reflectance begins by selecting the pixels in the image which contain the Spectralon reflectance target. Each pixel in this region contains 300 digital numbers; one digital number for each of the captured spectral channels. We then take the average value of all pixels within the region at each spectral channel, leaving us with a single, averaged digital number for each spectral channel. This process leaves us with a digital number that represents 99\% of the reflected light for each spectral channel. Finally, we calculate the spectral reflectance at each pixel as follows:
\[
\rho = \left(\frac{DN_{scene}-DN_{dark}}{DN_{target}-DN_{dark}}\right)\rho_{target},
    \label{eq:reflectance}
\]
where $\rho$ is the spectral reflectance, $DN_{scene}$ represents the digital number values captured in the image, $DN_{target}$ represents the averaged digital numbers of the reflectance target obtained through the process outlined above, $DN_{dark}$ represents the dark current or background signal generated through sporadic electron generation in the imager's sensor, and $\rho_{target}$ represents the reflectivity of the reflectance target.

We manually extracted 6,316 overlapping patches with a window size of $25 \times 25$ pixels from each of the 76 kochia images. Furthermore, we reduced the number of spectral bands within each patch from 300 to 150 by averaging adjacent pairs of bands, which can be interpreted as $2\times$ spectral binning, where the resulting spectral resolution of each channel has been modified from approximately \SI{2.12}{\nano\meter} to \SI{4.24}{\nano\meter}. As one of the goals of this work is to aid in the design of multispectral imaging systems, and it is unlikely that optical filters with a bandwidth less than \SI{20}{\nano\meter} will be used, decreasing the overall spectral resolution is unlikely to affect our results. Thus, this process itself gives us an upfront reduction in dimensionality that greatly reduces the potential overfitting impact in our following analysis.

Since the IP dataset consists of a single large image, we have to divide it into small patches so that each patch represents one class. Thus, we extracted square patches using a $5 \times 5$ pixel window around each pixel. Furthermore, we only collected patches around those pixels with an assigned label. By doing so, the new IP dataset has 10,249 patches.

Finally, we applied $z$-score normalization (mean equal to 0 and standard deviation equal to 1) onto each spectral band for both the Kochia and Indian Pines datasets.

\subsection{Inter-Band Redundancy Analysis} \label{preselection}

The first step of our method is to reduce the inter-band redundancy by selecting a subset of representative spectral bands. We utilized a filter-based selection method whereby we iteratively calculate the Variance Inflation Factor (VIF) \cite{StatLearn13} between pairs of bands in order to determine how correlated they are; that is, to verify the presence of collinearity between them. We call this Inter-Band Redundancy Analysis (IBRA).

The VIF value between two bands, $b_1$ and $b_2$, is calculated based on the R-squared value from the Ordinary Least Square (OLS) regression model built by taking one of the bands as a dependant variable ($b_1$) and the other as the independent variable ($b_2$). Specifically, $VIF(b_1,b_2) = 1/ (1- R^2_{b_1,b_2})$. A high VIF value means that the independent variable and the dependent variable explain the same variance within the dataset and are redundant. We will consider VIF values greater than a threshold $\theta$ as representing the presence of collinearity in the model. In the literature, the recommended values of $\theta$ are between 5 and 10 \cite{belsley}, so we test different values of $\theta \in [5, 12]$ to observe how the performance is affected and to choose the best $\theta$ for a given classification task.

While some methods, such as that proposed by \cite{castaldi}, use the VIF metric to identify and remove redundant spectral bands from a given set directly, our approach is novel and distinct in that we use it as part of a pre-selection step, assessing the collinearity degree between each band and its local neighbors iteratively in order to find independent salient bands. Thus, using the VIF metric, we calculate the number of bands $d_{left}(x)$ we need to move away to the left side from the $x$-th band in order to find bands sufficiently different from band $x$; similarly, we calculate the number of bands $d_{right}(x)$ we need to move away to the right side from the $x$-th band in order to find bands sufficiently different from that in band $x$ (see Algorithm~\ref{alg:vif}).
In this algorithm, we calculate the difference $d(x) = |d_{left}(x) - d_{right}(x)|$ for each spectral band to determine how this difference is distributed across the spectrum. Let $N$ be the total number of spectral bands within the dataset. Then {\tt getVIF($x, y$)} calculates the VIF value between bands at positions $x$ and $y$. We construct $\texttt{table}$, which is a symmetric matrix that stores the pre-computed VIF values between pairs of bands in order to avoid re-calculations. Then $\texttt{getLocalMinima}(x)$ is a function that retrieves the position of the local minimum points of the vector $x$. We consider only local minimum points where $d(x) < 5$; otherwise, they will not be suitable bandwidth centers. 

\begin{algorithm}[t!]
\scriptsize
\caption{Calculating the interband redundancy}
\begin{algorithmic}[1]
\Function{interbandRedundancy}{$\theta$}
    \State $d_{left} \leftarrow []$
    \State $d_{right} \leftarrow []$
    \State $table \leftarrow zeros(N,N)$ // creates an $N\times N$ matrix of zeros
    \ForAll {$band \in (0,N)$}
        \State {// Check left side}
        \State $t \leftarrow 1$    
        \State $vif \leftarrow \infty$
        \While {$(vif >\theta) \wedge ((band - dt) >0)$}
            \If {$table[band, band - t] = 0$}
                \State $table[band, band - t] = getVIF(band, band - t)$
                \State $table[band - t, band] = table[band, band - t]$
            \EndIf
            \State $vif = table[band, band - t]$
            \State $t \leftarrow t + 1$
        \EndWhile
        \State $d_{left} \leftarrow [d_{left} - 1]$
        \State{// Check right side}
        \State $t \leftarrow 1$    
        \State $vif \leftarrow \infty$
        \While {$(vif >\theta) \wedge ((band - dt) < N)$}
            \If {$table[band, band + t] = 0$}
                \State $table[band, band + t] = getVIF(band, band + t)$
                \State $table[band + t, band] = table[band, band + t]$
            \EndIf
            \State $vif = table[band, band + t]$
            \State $t \leftarrow t + 1$
        \EndWhile
        \State $d_{right} \leftarrow [d_{right} - 1]$
    \EndFor
    \State $d \leftarrow |d_{left} - d_{right}|$
    \State $preselection \leftarrow getLocalMinima(d)$
    \State \Return ${d, preselection}$
\EndFunction
\end{algorithmic}
\label{alg:vif}
\end{algorithm}

\begin{figure*}[t!]
\centering
\includegraphics[height = 4.8cm]{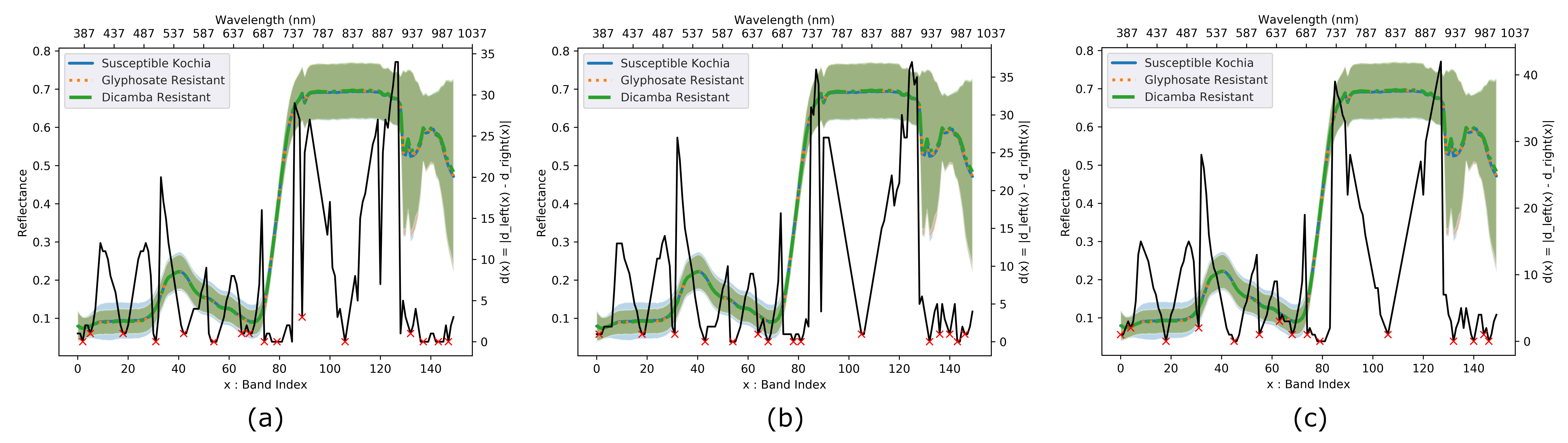}
\vspace{-2ex}
\caption{Spectral response and Spectral index vs. distance $d$ plots for the Kochia dataset using different VIF thresholds. \textbf{(a)} $\texttt{th=12}$ \textbf{(b)} $\texttt{th=10}$. \textbf{(c)} $\texttt{th=8}$. Local minima in the three graphs are indicated with an `$\times$'.}
\label{fig:dist}
\end{figure*}

The distribution of the variable $d(x)$ is used to find clusters with similar bands and their corresponding cluster centers. Since we are interested in choosing suitable bandwidth centers, we choose bands that are similar to both the left and right sides; that is, the difference $d(x)$ has to be minimum. Fig.~\ref{fig:dist} shows the distribution of the variable $d(x)$ for the Kochia dataset given different thresholds $\theta$. From this, we observe that each distribution consists of a series of ``V'' patterns. In this context, a local minimum---the center of a ``V'' pattern---represents a salient band that explains the variance within the dataset in a way similar to its neighbors. Even though all the bands within a ``V'' pattern are similar, the band at the leftmost position is more similar to those bands on the left side. Similarly, the band at the rightmost position of the ``V'' is more similar to those bands on the right side, while the band corresponding to the local minimum is similar to both sides, acting as a centroid. In general, we keep the bands corresponding to a local minimum in the plot of spectral index vs. $d(x)$ and remove the rest since they are redundant. 


\subsection{Band Selection Using Pre-Selected Bands} \label{sel}

In Sec.~\ref{preselection} we showed how to pick a set of independent candidate bands from the spectrum based on collinearity. Here, we employ a wrapper-based method to select the best combination of bands $S_f \in \mathbb{Z}^k$, given the desired number of bands $k$, from the set of available candidate bands $S_c \in \mathbb{Z}^{N'}$ (where $N'\ll N$ is the number of bands pre-selected by IBRA), which greatly reduces the search complexity. We call this process Greedy Spectral Selection (GSS). 

The first step is to rank each band $s_c \in S_c$ according to some criterion. In this work, we use information entropy to calculate an initial relevance score for each spectral band. Given a band $s_c$, which is considered a discrete random variable with a bit depth of 14 bits, its entropy $H(s_c)$ conveys the average level of information inherent in $s_c$ and is defined as follows:
\[
H(s_c) = -\sum_{z \in \Omega_{s_c}} P(z) \log P(z),
\]
where $P(z)$ is the probability mass function of random variable $z$, and $\Omega_{s_c}$ is the space that encompasses all the possible values that can occur in the spectral band $s_c$.

Other methods, such as those proposed by Wang et al. \cite{cluster,fast}, use information entropy to select the most representative band within each cluster. Our approach is different in the sense that we already picked the most relevant band from each cluster (defined as a local minimum in the plot of spectral index vs. distance $d(x)$); now, we rank the pre-selected bands based on information entropy in order to select a subset $S_f$ of $k$ bands. Thus, initially, $S_f$ consists of the top-$k$ bands of $S_c$ with the greatest entropy. 

Even if calculations show that collinearity does not exist between pairs of bands, it is possible that three or more bands are highly correlated---a phenomenon known as multicollinearity \cite{belsley}. In that case, we check for the presence of multicollinearity within the $k$ selected spectral bands the same way as described in Sec.~\ref{preselection} with collinearity.
With this, we employ the following greedy algorithm for band selection (see Algorithm~\ref{alg:selection}): First, we calculate the average classification performance (F1 score) when using the current $k$ selected bands; to do this, we perform a $5\times 2$ cross-validation and calculate the average F1 score obtained on the 10 validation folds. Then, after calculating the VIF value for each of the currently selected bands, we remove the one with the greatest VIF value and consider the next available band with the greatest entropy. With this new subset of $k$ bands, we train a new classifier and verify if the average classification performance has improved. We repeat this process until there are no more available bands or when we find a significant drop in performance. Finally, we select the combination of bands that showed the best classification performance. 

Note that Algorithm~\ref{alg:selection} selects $k$ band indices from a list of candidate band indices $S_c$. Function {\tt getEntropy($S_f$)} returns the entropy value of each of the candidate bands in $S_c$. Then, {\tt $S_c$.sort(key=$H$)} decreasingly sorts the elements of $S_c$ with respect to their corresponding entropy values. Function {\tt getVIFMulti($S_f$)} returns a list with the VIF value for each band in the list of selected bands $S_f$. Next {\tt getMax($l$)} returns the position where the maximum value in a list $l$ was found. Finally, {\tt trainSelection($S_f$)} returns the average F1 score evaluated using the bands in $S_f$ and $5\times 2$ cross-validation. 

\begin{algorithm} [t]
\scriptsize
\caption{Greedy spectral selection}
\begin{algorithmic}[1]
\Function{selectBands}{$S_c, k$}
    \State $H \leftarrow getEntropy(S_c)$
    \State $S_c.sort(key=H)$
    \State $S_f \leftarrow S_c[1:k]$
    \State $S_c \leftarrow S_c[k + 1: end]$
    \State $F1 \leftarrow trainSelection(S_f)$
    \State $bestS_f \leftarrow S_f$
    \While {$length(c) > 0$}
        \State $listVIF \leftarrow getVIFMulti(S_f)$ \label{line:vif}
        \State $index = getMax(listVIF)$
        \State $S_f[index : end] \leftarrow S_f[index + 1: end]$
        \State $S_f.append(S_c[1])$
        \State $S_c \leftarrow S_c[2 : end]$
        \State $newF1 \leftarrow trainSelection(S_f)$
        \If {$newF > F1$}
            \State $bestS_f \leftarrow S_f$
            \State $F1 \leftarrow newF1$
        \Else 
            \If {$newF1 <= F1 - 0.05$}
                \State break  // stop if a drop of 5\% is found
            \EndIf
        \EndIf
    \EndWhile
    \State \Return $bestS_f$
\EndFunction
\end{algorithmic}
\label{alg:selection}
\end{algorithm}


\subsection{Convolutional Neural Network Architecture} \label{cnn}

For all of our experiments, we used a modified version of the Hyper3DNet network \cite{hyper3dnet}, which is a 3D-2D CNN architecture specifically designed to solve HSI classification problems using a reduced number of trainable parameters. Furthermore, experimental results demonstrated relative superiority of this architecture over state-of-the-art architectures. 

In this paper, our modified network, referred to as Hyper3DNet-Lite, is a simplification of the original Hyper3DNet architecture. The difference with the original architecture is that its 3-D feature extractor consists of two 3-D convolutional layers instead of a densely connected block with four layers; additionally, its 2-D spatial encoder has three layers instead of four. The Hyper3DNet-Lite architecture used for the Kochia dataset is detailed in Table \ref{tab:k1}, where $N$ denotes the number of spectral bands in the input, ``SepConv2D" denotes a 2-D separable convolutional layer, and ``ReLU" denotes a rectified linear unit activation layer (where $ReLU(x)=\max (0, x))$. The only difference with the network used to process the IP dataset is that, since the input image is smaller ($5\times5$ pixels), the stride used in the last two ``SepConv2D" layers is $(1,1)$ instead of $(2,2)$ to avoid dimensionality inconsistencies.

The simplified architecture of the Hyper3DNet-Lite network becomes especially suitable for datasets that use just a few spectral bands, given that these datasets do not require models with a high level of complexity to process them, unlike those datasets that use all the available spectral bands. In this way, we avoid overparameterization, which results in our models being less prone to overfitting.

Previously, we also experimented with other types of classifiers (i.e. support vector machines, random forests, and feedforward neural networks) to use in the GSS process. However, due to the fast convergence rates and the substantial improvements on performance, we continued to use CNNs over the other types of classifiers.    

\setlength{\tabcolsep}{4pt}
\begin{table}
\begin{center}
\caption{Hyper3DNet-Lite architecture for the Kochia dataset.}
\resizebox{\columnwidth}{!}{
\label{tab:k1}

\small
\begin{tabular}{|c|c|c|c|}
\hline
\textbf{Layer Name}   & \textbf{Kernel Size} & \textbf{Stride Size} & \textbf{Output Size} \\ \hline
Input                 & ---                    & ---                    & (25, 25, $N$, 1)     \\ \hline
Conv3D + ReLU         & (3, 3, 3)          & (1, 1, 1)            & (25, 25, $N$, 16)     \\ \hline
Conv3D + ReLU         & (3, 3, 3)          & (1, 1, 1)            & (25, 25, $N$, 16)     \\ \hline
Reshape               & ---                    & ---                    & (25, 25, $16N$)      \\ \hline
SepConv2D + ReLU & (3, 3)               & (1, 1)               & (25, 25, 320)        \\ \hline
SepConv2D + ReLU & (3, 3)               & (2, 2)               & (13, 13, 256)        \\ \hline
SepConv2D + ReLU & (3, 3)               & (2, 2)               & (7, 7, 256)          \\ \hline
GlobalAveragePooling  & ---                    & ---                    & 256                 \\ \hline
Dense + Softmax       & ---                    & ---                    & \# classes                    \\ \hline
\end{tabular}
}
\end{center}
\vspace{-2ex}
\end{table}


\subsection{Multispectral Sensor Design}

The previous steps are used to select the most relevant spectral bands from the original hyperspectral data cube. However, knowing which wavelengths are the most useful for a given application allows for the design of compact multispectral sensors instead of using a full hyperspectral sensor. To accomplish this, we use the original data cube and simulate applying optical filters to capture data from a multispectral imager.

To do this, we generate $k$ Gaussian distributions, taking the position of the spectral bands selected by the GSS method as the centroids. The bandwidth of these distributions is set to five bands or, equivalently, \SI{20}{\nano\meter}, to represent a common optical filter bandwidth. The simulated multispectral reflectance measurement is obtained by multiplying the original hyperspectral data cube by the corresponding Gaussian distribution generated for each band, then integrating under the resulting response curve to get a single reflectance value. This process is repeated for each of the $k$ Gaussian distributions.


\section{Experimental Results}

For the sake of consistency and fairness, we used the same configuration (i.e., network architecture, optimizer, and batch size) for all the networks trained in our experiments. While this strategy does not guarantee the best possible results, it allows us to compare the behavior of different band selection methods under the same conditions. All CNNs were trained using the Adadelta optimizer \cite{adadelta}, which is a gradient descent method based on an adaptive learning rate, so that there is no need to select a global learning rate manually. The mini-batch size was set to 128. The last layer of the CNNs used a softmax activation function, and we employed a categorical cross-entropy loss function. Furthermore, we used $5\times 2$-fold stratified cross-validation to train and evaluate all networks. Note that $z$-score normalization was applied to each training set while the exact same scaling was applied to their corresponding validation set. In order to analyze the behavior of our models, we calculated four metrics on the validation sets: accuracy ($OA$), macro-average precision ($Prec$), macro-average recall ($Rec$), and F1 score. The source code and datasets are available online\footnote{Codebase: \url{https://github.com/GiorgioMorales/HSI-BandSelection.git}.}.

In the following sections, we compare the results of using our inter-band redundancy analysis strategy alone and our greedy spectral selection strategy after pre-selection. We also compare our results with several state-of-the-art methods for bandwidth selection. For all our experiments, we select a reduced number of spectral bands $k$, as our objective is to design simple task-specific multispectral sensors.

\subsection{Training Pre-Selected Bands} \label{tpreselection}

Previously (Fig.~\ref{fig:dist}) we showed some examples of applying the pre-selection method using IBRA on the Kochia dataset using three different VIF thresholds (12, 10, and 8), which reduced our search space from 150 bands to 19, 17, and 16 bands, respectively. Table~\ref{tab:base} gives the number of pre-selected bands for both the Kochia and IP datasets when using a VIF threshold of 10; it also gives the average performance for the four metrics and corresponding standard deviations using the Hyper3DNet-Lite network when training on the full hyperspectral spectrum and only the pre-selected bands. The number of parameters required to train each network is reported in the last column.

\setlength{\tabcolsep}{4pt}
\begin{table}[t]
\begin{center}
\caption{Performance with and without IBRA preselection ($\theta=10$).}
\resizebox{\columnwidth}{!}{
\label{tab:base}
\begin{tabular}{|c |c |c|c|c|c|c|}
\hline\noalign{}
 \textbf{Dataset} & \textbf{\# Bands} & \textbf{OA}  & \textbf{Prec}  & \textbf{Rec} & \textbf{F1} & \textbf{\# Param.} \\ \hline

 \multirow{3}{*}{\textbf{Kochia}} & 150 &
\begin{tabular}[c]{@{}c@{}}98.46 \\ $\pm$ 0.29\end{tabular} & 
\begin{tabular}[c]{@{}c@{}}98.66 \\ $\pm$ 0.26\end{tabular} & 
\begin{tabular}[c]{@{}c@{}}98.55 \\ $\pm$ 0.31\end{tabular} & 
\begin{tabular}[c]{@{}c@{}}98.60 \\ $\pm$ 0.28\end{tabular} & 561,475\\ \cline{2-7} &
 17 &
\begin{tabular}[c]{@{}c@{}}97.05 \\ $\pm$ 0.47\end{tabular} & 
\begin{tabular}[c]{@{}c@{}}97.25 \\ $\pm$ 0.45\end{tabular} & 
\begin{tabular}[c]{@{}c@{}}97.17 \\ $\pm$ 0.46\end{tabular} & 
\begin{tabular}[c]{@{}c@{}}97.21 \\ $\pm$ 0.44\end{tabular} & 258,035  \\ \cline{2-7} 
 \hline
 
 \multirow{3}{*}{\textbf{Indian Pines}} & 200 &
\begin{tabular}[c]{@{}c@{}}99.42 \\ $\pm$ 0.18\end{tabular} & 
\begin{tabular}[c]{@{}c@{}}99.32 \\ $\pm$ 0.29\end{tabular} & 
\begin{tabular}[c]{@{}c@{}}99.47 \\ $\pm$ 0.28\end{tabular} & 
\begin{tabular}[c]{@{}c@{}}99.39 \\ $\pm$ 0.27\end{tabular} & 1,274,464\\ \cline{2-7} &
 31  &
 \begin{tabular}[c]{@{}c@{}}99.49 \\ $\pm$ 0.14\end{tabular} & \begin{tabular}[c]{@{}c@{}}99.38 \\ $\pm$ 0.34\end{tabular} & \begin{tabular}[c]{@{}c@{}}99.56 \\ $\pm$ 0.19\end{tabular} & \begin{tabular}[c]{@{}c@{}}99.47 \\ $\pm$ 0.23\end{tabular} & 338,880\\ \cline{2-7} 
 
 \hline
\end{tabular}
}
\end{center}
\end{table}

\subsection{Greedy Spectral Selection} \label{GSS}

\setlength{\tabcolsep}{4pt}
\begin{table}[t]
\centering
\begin{center}
\caption{Greedy Spectral Selection on the Kochia dataset.}
\resizebox{\columnwidth}{!}{
\label{tab:GSSKochia}
 \def\arraystretch{1.2}%
\begin{tabular}{|l|c|c|c|c|c|c|}
\hline
\textbf{\textit{k}} & \textbf{VIF} & \textbf{Selected bands (\SI{}{\nano\meter})} & \textbf{OA} & \textbf{Prec} & \textbf{Rec} & \textbf{F1} \\ \hline
 & 12 & \begin{tabular}[c]{@{}c@{}}{[}395.5, 463.3, 565.1,\\ 700.8, 722.0 , 993.3{]}\end{tabular} & \begin{tabular}[c]{@{}c@{}}92.44 \\ $\pm$ 0.71\end{tabular} & \begin{tabular}[c]{@{}c@{}}92.76 \\ $\pm$ 0.80\end{tabular} & \begin{tabular}[c]{@{}c@{}}92.79 \\ $\pm$ 0.67\end{tabular} & \begin{tabular}[c]{@{}c@{}}92.76 \\ $\pm$ 0.72\end{tabular} \\ \cline{2-7} 
 & 11 & \begin{tabular}[c]{@{}c@{}}{[}395.5, 408.2, 463.3, \\586.3, 662.6, 700.8{]}\end{tabular} & \begin{tabular}[c]{@{}c@{}}90.74 \\ $\pm$ 1.05\end{tabular} & \begin{tabular}[c]{@{}c@{}}91.56 \\ $\pm$ 0.97\end{tabular} & \begin{tabular}[c]{@{}c@{}}91.54 \\ $\pm$ 1.06\end{tabular} & \begin{tabular}[c]{@{}c@{}}91.54 \\ $\pm$ 1.01\end{tabular} \\ \cline{2-7} 
 & \textbf{10} & \textbf{\begin{tabular}[c]{@{}c@{}}{[}391.2, 463.3, 569.3, \\675.3, 730.4, 993.3{]}\end{tabular}} & 
 \begin{tabular}[c]{@{}c@{}}92.69 \\ $\pm$ 0.53\end{tabular} &
 \begin{tabular}[c]{@{}c@{}}93.24 \\ $\pm$ 0.52\end{tabular} &
 \begin{tabular}[c]{@{}c@{}}93.08 \\ $\pm$ 0.49\end{tabular} &
 \textbf{\begin{tabular}[c]{@{}c@{}}93.15 \\ $\pm$ 0.49\end{tabular}} \\ \cline{2-7} 
 & 9 & \begin{tabular}[c]{@{}c@{}}{[}391.2, 463.3, 569.3, \\700.8, 730.4, 993.3{]}\end{tabular} & \begin{tabular}[c]{@{}c@{}}92.40 \\ $\pm$ 0.63\end{tabular} & \begin{tabular}[c]{@{}c@{}}92.67 \\ $\pm$ 0.63\end{tabular} & \begin{tabular}[c]{@{}c@{}}92.77 \\ $\pm$ 0.59\end{tabular} & \begin{tabular}[c]{@{}c@{}}92.71 \\ $\pm$ 0.59\end{tabular} \\ \cline{2-7} 
 & 8 & \begin{tabular}[c]{@{}c@{}}{[}387.0 , 404.0 , 463.3,\\ 577.8, 700.8, 722.0{]}\end{tabular} & \begin{tabular}[c]{@{}c@{}}92.58 \\ $\pm$ 0.63\end{tabular} & \begin{tabular}[c]{@{}c@{}}93.05 \\ $\pm$ 0.65\end{tabular} & \begin{tabular}[c]{@{}c@{}}93.08 \\ $\pm$ 0.57\end{tabular} & \begin{tabular}[c]{@{}c@{}}93.06 \\ $\pm$ 0.59\end{tabular} \\ \cline{2-7} 
 & 7 & \begin{tabular}[c]{@{}c@{}}{[}387.0, 404.0, 463.3, \\569.3, 700.8, 722.0{]}\end{tabular} & \begin{tabular}[c]{@{}c@{}}92.07 \\ $\pm$ 0.89\end{tabular} & \begin{tabular}[c]{@{}c@{}}92.52 \\ $\pm$ 0.89\end{tabular} & \begin{tabular}[c]{@{}c@{}}92.55 \\ $\pm$ 0.79\end{tabular} & \begin{tabular}[c]{@{}c@{}}92.53 \\ $\pm$ 0.83\end{tabular} \\ \cline{2-7} 
 & 6 & \begin{tabular}[c]{@{}c@{}}{[}387.0 , 404.0 , 463.3, \\586.3, 700.8, 717.7{]}\end{tabular} & \begin{tabular}[c]{@{}c@{}}92.00 \\ $\pm$ 0.61\end{tabular} & \begin{tabular}[c]{@{}c@{}}92.57 \\ $\pm$ 0.54\end{tabular} & \begin{tabular}[c]{@{}c@{}}92.52 \\ $\pm$ 0.64\end{tabular} & \begin{tabular}[c]{@{}c@{}}92.53 \\ $\pm$ 0.57\end{tabular} \\ \cline{2-7} 
\multirow{-16}{*}{6} & 5 & {\begin{tabular}[c]{@{}c@{}}{[}387.0 , 463.3, 586.3, \\645.6, 700.8, 722.0{]}\end{tabular} } &
\begin{tabular}[c]{@{}c@{}}91.03 \\ $\pm$ 1.04\end{tabular} & \begin{tabular}[c]{@{}c@{}}91.79 \\ $\pm$ 1.14\end{tabular} & \begin{tabular}[c]{@{}c@{}}91.75 \\ $\pm$ 0.91\end{tabular} & \begin{tabular}[c]{@{}c@{}}91.76 \\ $\pm$ 1.01\end{tabular} \\ \hline

 & 12 & \begin{tabular}[c]{@{}c@{}}{[}395.5, 408.2 , 463.3, 518.4, 565.1, \\ 616.0, 675.3, 700.8, 722.0, 993.3{]}\end{tabular} & \begin{tabular}[c]{@{}c@{}}96.31\\ $\pm$ 0.69\end{tabular} & \begin{tabular}[c]{@{}c@{}}96.57\\ $\pm$ 0.55\end{tabular} & \begin{tabular}[c]{@{}c@{}}96.49 \\ $\pm$ 0.73\end{tabular} & \begin{tabular}[c]{@{}c@{}}96.53 \\ $\pm$ 0.64\end{tabular} \\ \cline{2-7} 
 & 11 & \begin{tabular}[c]{@{}c@{}}{[}395.5, 408.2, 463.3, 565.1, 662.6, \\675.3, 700.8, 713.5, 726.2, 993.3{]}\end{tabular} & \begin{tabular}[c]{@{}c@{}}96.18 \\ $\pm$ 0.41\end{tabular} & \begin{tabular}[c]{@{}c@{}}96.48\\ $\pm$ 0.29\end{tabular} & \begin{tabular}[c]{@{}c@{}}96.31 \\ $\pm$ 0.46\end{tabular} & \begin{tabular}[c]{@{}c@{}}96.39\\ $\pm$ 0.36\end{tabular} \\ \cline{2-7} 
 & 10 & {\begin{tabular}[c]{@{}c@{}}{[}391.2, 463.3, 518.4, 569.3,  658.4, \\675.3, 717.7, 730.4, 993.3, 1006.{]}\end{tabular}} & \begin{tabular}[c]{@{}c@{}}95.83\\ $\pm$ 0.36\end{tabular} & \begin{tabular}[c]{@{}c@{}}96.10\\ $\pm$ 0.38\end{tabular} & \begin{tabular}[c]{@{}c@{}}96.06\\ $\pm$ 0.32\end{tabular} & \begin{tabular}[c]{@{}c@{}}96.08\\ $\pm$ 0.34\end{tabular} \\ \cline{2-7} 
 & 9 & \begin{tabular}[c]{@{}c@{}}{[}391.2, 463.3, 518.4, 569.3, 616.0, \\671.1, 700.8, 717.7, 730.4, 993.3{]}\end{tabular} & \begin{tabular}[c]{@{}c@{}}96.16\\ $\pm$ 0.56\end{tabular} & \begin{tabular}[c]{@{}c@{}}96.48\\ $\pm$ 0.50\end{tabular} & \begin{tabular}[c]{@{}c@{}}96.37 \\ $\pm$ 0.54\end{tabular} & \begin{tabular}[c]{@{}c@{}}96.42\\ $\pm$ 0.52\end{tabular} \\ \cline{2-7} 
 & 8 & \begin{tabular}[c]{@{}c@{}}{[}387.0, 404.0, 463.3, 518.4,  577.8, \\654.1, 675.3, 700.8, 722.0, 1006.0{]}\end{tabular} & \begin{tabular}[c]{@{}c@{}}96.47\\ $\pm$ 0.38\end{tabular} & \begin{tabular}[c]{@{}c@{}}96.79\\ $\pm$ 0.36\end{tabular} & \begin{tabular}[c]{@{}c@{}}96.66\\ $\pm$ 0.37\end{tabular} & \begin{tabular}[c]{@{}c@{}}96.72\\ $\pm$ 0.35\end{tabular} \\ \cline{2-7} 
 & \textbf{7} & \textbf{\begin{tabular}[c]{@{}c@{}}{[}387.0, 404.0,  463.3, 518.4, 569.3, \\654.1, 675.3, 700.8, 722.0, 1006.0{]}\end{tabular}} & 
 \begin{tabular}[c]{@{}c@{}}96.69\\ $\pm$ 0.35\end{tabular} &
 \begin{tabular}[c]{@{}c@{}}96.92\\ $\pm$ 0.38\end{tabular} &
 \begin{tabular}[c]{@{}c@{}}96.95\\ $\pm$ 0.34\end{tabular} &
 \textbf{\begin{tabular}[c]{@{}c@{}}96.93\\ $\pm$ 0.35\end{tabular}} \\ \cline{2-7} 
 & 6 & {\begin{tabular}[c]{@{}c@{}}{[}387.0, 404.0, 463.3, 586.3,  649.9, \\679.6, 700.8, 717.7, 730.4, 1001.8{]}\end{tabular}} & \begin{tabular}[c]{@{}c@{}}95.91\\ $\pm$ 0.50\end{tabular} & \begin{tabular}[c]{@{}c@{}}96.34\\ $\pm$ 0.44\end{tabular} & \begin{tabular}[c]{@{}c@{}}96.12\\ $\pm$ 0.47\end{tabular} & \begin{tabular}[c]{@{}c@{}}96.22\\ $\pm$ 0.45\end{tabular} \\ \cline{2-7} 
\multirow{-16}{*}{10} & 5 & {\begin{tabular}[c]{@{}c@{}}{[}387.0,  463.3, 586.3, 645.6, 700.8,\\ 722.0, 832.2, 946.7, 980.6, 1001.8{]}\end{tabular}} & { \begin{tabular}[c]{@{}c@{}}95.06\\ $\pm$ 0.54\end{tabular}} & \begin{tabular}[c]{@{}c@{}}95.44\\ $\pm$ 0.52\end{tabular} & \begin{tabular}[c]{@{}c@{}}95.33\\ $\pm$ 0.56\end{tabular} & \begin{tabular}[c]{@{}c@{}}95.38\\ $\pm$ 0.53\end{tabular} \\ \hline
\end{tabular}
}
\end{center}
\end{table}

 \setlength{\tabcolsep}{4pt}
\begin{table}[!ht]
\centering
\begin{center}
\caption{Greedy Spectral Selection on the Indian Pines dataset.}
\resizebox{\columnwidth}{!}{
\label{tab:GSSIP}
 \def\arraystretch{1.2}%
\begin{tabular}{|c|c|c|c|c|c|}
\hline
\textbf{VIF} & \textbf{Selected bands (\SI{}{\nano\meter})} & \textbf{OA} & \textbf{Prec} & \textbf{Rec} & \textbf{F1} \\ \hline
 12 & \begin{tabular}[c]{@{}c@{}}{[}484.6,  627.2,  703.3, 750.8, 1017.1{]}\end{tabular} & \begin{tabular}[c]{@{}c@{}}97.96\\ $\pm$ 0.33\end{tabular} & \begin{tabular}[c]{@{}c@{}}98.21\\ $\pm$ 0.43\end{tabular} & \begin{tabular}[c]{@{}c@{}}98.32\\ $\pm$ 0.33\end{tabular} & \begin{tabular}[c]{@{}c@{}}98.25\\ $\pm$ 0.35\end{tabular} \\ 
 \hline
 11 & \begin{tabular}[c]{@{}c@{}}{[}541.7,  570.2,  703.3, 750.8, 1017.1{]}\end{tabular} & \begin{tabular}[c]{@{}c@{}}97.55\\ $\pm$ 0.29\end{tabular} & \begin{tabular}[c]{@{}c@{}}98.05\\ $\pm$ 0.29\end{tabular} & \begin{tabular}[c]{@{}c@{}}97.95\\ $\pm$ 0.29\end{tabular} & \begin{tabular}[c]{@{}c@{}}97.98\\ $\pm$ 0.22\end{tabular} \\ 
 \hline
 \textbf{10} & { \textbf{\begin{tabular}[c]{@{}c@{}}{[}484.6,  617.7,  703.3, 750.8, 1017.1{]}\end{tabular}}} & 
 \begin{tabular}[c]{@{}c@{}}98.08\\ $\pm$ 0.43\end{tabular} & 
 \begin{tabular}[c]{@{}c@{}}98.26\\ $\pm$ 0.42\end{tabular} & 
 \begin{tabular}[c]{@{}c@{}}98.39\\ $\pm$ 0.43\end{tabular} & \textbf{\begin{tabular}[c]{@{}c@{}}98.32\\ $\pm$ 0.39\end{tabular}} \\ 
 \hline
 9 & \begin{tabular}[c]{@{}c@{}}{[}541.7,  617.7,  703.3, 817.4, 1017.1{]}\end{tabular} & \begin{tabular}[c]{@{}c@{}}98.28\\ $\pm$ 0.35\end{tabular} & \begin{tabular}[c]{@{}c@{}}98.24\\ $\pm$ 0.47\end{tabular} & \begin{tabular}[c]{@{}c@{}}98.06\\ $\pm$ 0.59\end{tabular} & \begin{tabular}[c]{@{}c@{}}98.11\\ $\pm$ 0.43\end{tabular} \\ 
 \hline 
 8 & \begin{tabular}[c]{@{}c@{}}{[}589.2,  627.2,  703.3,  817.4, 1017.1{]}\end{tabular} & \begin{tabular}[c]{@{}c@{}}98.04\\ $\pm$ 0.30\end{tabular} & \begin{tabular}[c]{@{}c@{}}98.19\\ $\pm$ 0.46\end{tabular} & \begin{tabular}[c]{@{}c@{}}98.06\\ $\pm$ 0.46\end{tabular} & \begin{tabular}[c]{@{}c@{}}98.10\\ $\pm$ 0.35\end{tabular} \\ 
 \hline
 7 & \begin{tabular}[c]{@{}c@{}}{[}551.2,  570.2,  703.3,  817.4, 1017.1{]}\end{tabular} & \begin{tabular}[c]{@{}c@{}}98.01\\ $\pm$ 0.24\end{tabular} & \begin{tabular}[c]{@{}c@{}}98.29\\ $\pm$ 0.18\end{tabular} & \begin{tabular}[c]{@{}c@{}}98.36\\ $\pm$ 0.42\end{tabular} & \begin{tabular}[c]{@{}c@{}}98.31\\ $\pm$ 0.26\end{tabular} \\ 
 \hline
 6 & \begin{tabular}[c]{@{}c@{}}{[}560.7,  703.3,  817.4,  912.5, 1017.1{]}\end{tabular}  & \begin{tabular}[c]{@{}c@{}}97.06\\ $\pm$ 0.49\end{tabular} & \begin{tabular}[c]{@{}c@{}}97.49\\ $\pm$ 0.58\end{tabular} & \begin{tabular}[c]{@{}c@{}}97.63\\ $\pm$ 0.48\end{tabular} & \begin{tabular}[c]{@{}c@{}}97.53\\ $\pm$ 0.48\end{tabular} \\ 
 \hline
 5 & \begin{tabular}[c]{@{}c@{}}{[}560.7,  712.8,  807.9,  912.5, 1017.1{]}\end{tabular}  & \begin{tabular}[c]{@{}c@{}}96.73\\ $\pm$ 0.55\end{tabular} & \begin{tabular}[c]{@{}c@{}}97.46\\ $\pm$ 0.53\end{tabular} & \begin{tabular}[c]{@{}c@{}}97.00\\ $\pm$ 0.61\end{tabular} & \begin{tabular}[c]{@{}c@{}}97.19\\ $\pm$ 0.49\end{tabular} \\ \hline 
\end{tabular}
}
\end{center}
\end{table}

Next, we applied the GSS method for each of the sets of IBRA-selected bands using different VIF thresholds $\theta\in [5, 12]$. Then, we selected the classifier that achieved the best classification performance based on the mean F1-score obtained after a $5\times2$-fold cross-validation.
 
For the Kochia dataset, we considered six and ten bands in order to evaluate the trade-off between the number of bands and performance. For the IP dataset, we selected only five bands. In addition, for each dataset, we experimented with different dataset sizes (i.e. 100\%, 75\%, 50\%, and 25\%) to evaluate how consistent the band selection results are under different data set sizes.
 
For the Kochia dataset, when selecting six bands, the best results were obtained using a VIF threshold of ten ($\theta=10$) and the wavelengths of the selected bands (in \SI{}{\nano\meter}) were $[391.2, 463.3, 569.3, 675.3, 730.4, 993.3]$ 
for each of the four dataset size variations. When selecting ten bands, the best results were obtained using a VIF threshold of $\theta=7$ when using 100\% and 50\% of the dataset, and $\theta=8$ when using 75\% and 25\% of the dataset. The wavelengths of the bands selected for $\theta=7$ were $[387.0, 404.0, 463.3, 518.4, 569.3, 654.1, 675.3, 700.8, 722.0,$ $1006.0] \;$ 
and the only difference with respect to the bands selected for $\theta=8$ was the selection of the wavelength \SI{577.8}{\nano\meter} instead of \SI{569.3}{\nano\meter}. Table~\ref{tab:GSSKochia} shows the performance using IBRA and GSS on the full Kochia dataset, where the bold entries represent the best VIF threshold, band selection, and average F1 score.
 
For the IP dataset, the best results were obtained using a VIF threshold of ten ($\theta=10$) and the wavelength of the selected bands were $[484.6,  617.7,  703.3, 817.4, 1017.1]$ for all the four dataset size variations. Table~\ref{tab:GSSIP} shows the performance using IBRA and GSS on the full IP dataset.


\subsection{Comparative Results} \label{compresult}

Finally, we compared our IBRA-GSS method to three other methods: OCF \cite{cluster}, HAGRID \cite{Neil}, and PLS-DA \cite{PLSDA}. For OCF, we used the normalized cut objective function along with information entropy ranking, as they showed the best performance. For HAGRID, we used a grid search to choose the following hyperparameters: a crossover rate of 0.25, a mutation rate of 0.05, a tournament size of 5, a population size of 1,000, and 300 iterations. To analyze the effectiveness of the feature selection methods, we compared the performance of four CNNs, each trained on the features selected by the four methods. This comparison was carried out using the same network architecture, hyperparameters, and other configurations for all of the methods. Finally, to determine if the difference in performance scores was statistically significant, we performed a paired $t$-test using the F1 scores at the $\alpha = 0.05$ level.

The method comparison is shown in Table \ref{tab:rkochia} for the Kochia dataset and in Table \ref{tab:rIP} for the IP dataset, with the best performing metrics highlighted in bold. Here, the first row of each method represents the results obtained after training a model using the original selected bands (identified as ``original band selection"), while the second row represents the results obtained after using simulated filters that take the position of the selected bands as their central wavelengths (identified as ``multispectral filter simulation"). The simulated filters used for the Kochia dataset were \SI{20}{\nano\meter} while for the IP dataset were \SI{50}{\nano\meter}. According to the $t$-test, our method performed significantly better than the other four methods in each of the cases. Although not shown due to space limitations, we also tested the four dataset size variations, as explained previously, and found that the improvement in performance of our method over the others was still statistically significant even when the dataset size was reduced. Additional experiments with other values of $k$ showed that the improvements of GSS over the compared methods remained consistent.


\setlength{\tabcolsep}{4pt}
\begin{table}[t]
\small
\begin{center}
\caption{Feature selection method comparison --- Kochia.}
\small
\resizebox{\columnwidth}{!}{
\label{tab:rkochia}
 
\begin{tabular}{|c|c|c|c|c|c|c|c|c|c|c|c|c|}
\hline \noalign{}
\multicolumn{1}{|c|}{
\rule[-1ex]{0pt}{3.5ex}\textbf{Bands}}                             
& \multicolumn{4}{c|}{\textbf{6}}  
& \multicolumn{4}{c|}{\textbf{10}} \\ \hline

\multicolumn{1}{|c|}{
\rule[-1ex]{0pt}{3.5ex}\textbf{Method}}                           
& \textbf{OA}
& \textbf{Prec}                                             
& \textbf{Rec}                                             
& \textbf{F1}                                              
& \textbf{OA}                                               
& \textbf{Prec}                                            
& \textbf{Rec}                                             
& \textbf{F1}\\ \hline

\multirow{3}{*}{\textbf{FNGBS}}                               &  
\begin{tabular}[c]{@{}c@{}}84.32\\ $\pm$ 1.78\end{tabular}  &
\begin{tabular}[c]{@{}c@{}}84.85\\ $\pm$ 1.77\end{tabular}  &
\begin{tabular}[c]{@{}c@{}}84.37\\ $\pm$ 1.72\end{tabular}  &
\begin{tabular}[c]{@{}c@{}}84.59\\ $\pm$ 1.73\end{tabular}  &  
\begin{tabular}[c]{@{}c@{}}93.78\\ $\pm$  0.77\end{tabular}  &
\begin{tabular}[c]{@{}c@{}}94.17\\ $\pm$ 0.84\end{tabular}  &
\begin{tabular}[c]{@{}c@{}}93.99\\ $\pm$ 0.73\end{tabular}  &
\begin{tabular}[c]{@{}c@{}}94.08\\ $\pm$ 0.78\end{tabular}  \\ \cline{2-9} &
\begin{tabular}[c]{@{}c@{}}86.98\\ $\pm$ 0.84\end{tabular}  &
\begin{tabular}[c]{@{}c@{}}87.35\\ $\pm$ 0.80\end{tabular}  &
\begin{tabular}[c]{@{}c@{}}86.91\\ $\pm$ 0.91\end{tabular}  &
\begin{tabular}[c]{@{}c@{}}87.10\\ $\pm$ 0.83\end{tabular}  &
\begin{tabular}[c]{@{}c@{}}94.19\\ $\pm$ 0.47\end{tabular}  & 
\begin{tabular}[c]{@{}c@{}}94.54\\ $\pm$ 0.47\end{tabular}  &
\begin{tabular}[c]{@{}c@{}}94.27\\ $\pm$ 0.51\end{tabular}  &
\begin{tabular}[c]{@{}c@{}}94.39\\ $\pm$ 0.48\end{tabular}  \\ \hline

\multirow{3}{*}{\textbf{PLS-DA}}              & 
\begin{tabular}[c]{@{}c@{}}84.77\\ $\pm$ 1.83\end{tabular}  &
\begin{tabular}[c]{@{}c@{}}85.15\\ $\pm$ 1.89\end{tabular}  &
\begin{tabular}[c]{@{}c@{}}84.69\\ $\pm$ 1.82\end{tabular}  &
\begin{tabular}[c]{@{}c@{}}84.89\\ $\pm$ 1.82\end{tabular}  &
\begin{tabular}[c]{@{}c@{}}94.36\\ $\pm$ 0.51\end{tabular}  & 
\begin{tabular}[c]{@{}c@{}}94.86\\ $\pm$ 0.55\end{tabular}  &
\begin{tabular}[c]{@{}c@{}}94.67\\ $\pm$ 0.47\end{tabular}  &
\begin{tabular}[c]{@{}c@{}}94.76\\ $\pm$ 0.49\end{tabular}  \\ \cline{2-9} &
\begin{tabular}[c]{@{}c@{}}88.41\\ $\pm$ 0.79\end{tabular}  &
\begin{tabular}[c]{@{}c@{}}88.85\\ $\pm$ 0.62\end{tabular}  &
\begin{tabular}[c]{@{}c@{}}88.37\\ $\pm$ 0.96\end{tabular}  &
\begin{tabular}[c]{@{}c@{}}88.59\\ $\pm$ 0.78\end{tabular}  &
\begin{tabular}[c]{@{}c@{}}95.10\\ $\pm$ 0.68\end{tabular}  & 
\begin{tabular}[c]{@{}c@{}}95.44\\ $\pm$ 0.59\end{tabular}  &
\begin{tabular}[c]{@{}c@{}}95.18\\ $\pm$ 0.67\end{tabular}  &
\begin{tabular}[c]{@{}c@{}}95.30\\ $\pm$ 0.63\end{tabular}  \\ \hline

\multirow{3}{*}{\textbf{OCF}}                               &  
\begin{tabular}[c]{@{}c@{}}90.48\\ $\pm$ 0.57\end{tabular}  &
\begin{tabular}[c]{@{}c@{}}90.92\\ $\pm$ 0.62\end{tabular}  &
\begin{tabular}[c]{@{}c@{}}90.81\\ $\pm$ 0.44\end{tabular}  &
\begin{tabular}[c]{@{}c@{}}90.86\\ $\pm$ 0.49\end{tabular}  &
\begin{tabular}[c]{@{}c@{}}94.87\\ $\pm$ 0.51\end{tabular}  & 
\begin{tabular}[c]{@{}c@{}}95.23\\ $\pm$ 0.52\end{tabular}  &
\begin{tabular}[c]{@{}c@{}}95.11\\ $\pm$ 0.46\end{tabular}  &
\begin{tabular}[c]{@{}c@{}}95.16\\ $\pm$ 0.47\end{tabular}  \\ \cline{2-9} &
\begin{tabular}[c]{@{}c@{}}92.42\\ $\pm$ 0.67\end{tabular}  &
\begin{tabular}[c]{@{}c@{}}92.75\\ $\pm$ 0.66\end{tabular}  &
\begin{tabular}[c]{@{}c@{}}92.66\\ $\pm$ 0.66\end{tabular}  &
\begin{tabular}[c]{@{}c@{}}92.70\\ $\pm$ 0.65\end{tabular}  &
\begin{tabular}[c]{@{}c@{}}94.62\\ $\pm$ 0.73\end{tabular}  & 
\begin{tabular}[c]{@{}c@{}}95.00\\ $\pm$ 0.65\end{tabular}  &
\begin{tabular}[c]{@{}c@{}}94.80\\ $\pm$ 0.64\end{tabular}  &
\begin{tabular}[c]{@{}c@{}}94.89\\ $\pm$ 0.64\end{tabular}  \\ \hline

\multirow{3}{*}{\textbf{HAGRID}}                            &  
\begin{tabular}[c]{@{}c@{}}91.71\\ $\pm$ 0.83\end{tabular}  &
\begin{tabular}[c]{@{}c@{}}92.25\\ $\pm$ 0.78\end{tabular}  &
\begin{tabular}[c]{@{}c@{}}92.17\\ $\pm$ 0.84\end{tabular}  &
\begin{tabular}[c]{@{}c@{}}92.20\\ $\pm$ 0.80\end{tabular}  &
\begin{tabular}[c]{@{}c@{}}94.50\\ $\pm$ 0.81\end{tabular}  &
\begin{tabular}[c]{@{}c@{}}94.81\\ $\pm$ 0.78\end{tabular}  & 
\begin{tabular}[c]{@{}c@{}}94.69\\ $\pm$ 0.72\end{tabular}  &
\begin{tabular}[c]{@{}c@{}}94.74\\ $\pm$ 0.74\end{tabular}  \\ \cline{2-9} &
\begin{tabular}[c]{@{}c@{}}92.48\\ $\pm$ 0.62\end{tabular}  &
\begin{tabular}[c]{@{}c@{}}92.91\\ $\pm$ 0.53\end{tabular}  &
\begin{tabular}[c]{@{}c@{}}92.89\\ $\pm$ 0.58\end{tabular}  &
\begin{tabular}[c]{@{}c@{}}92.89\\ $\pm$ 0.54\end{tabular}  &
\begin{tabular}[c]{@{}c@{}}95.14\\ $\pm$ 0.51\end{tabular}  & 
\begin{tabular}[c]{@{}c@{}}95.49\\ $\pm$ 0.48\end{tabular}  &
\begin{tabular}[c]{@{}c@{}}95.18\\ $\pm$ 0.51\end{tabular}  &
\begin{tabular}[c]{@{}c@{}}95.33\\ $\pm$ 0.47\end{tabular}  \\ \hline

\multirow{3}{*}{\textbf{GSS}}                               & 
\begin{tabular}[c]{@{}c@{}}\textbf{92.69}\\ \textbf{$\pm$ 0.53}\end{tabular}  & 
\begin{tabular}[c]{@{}c@{}}\textbf{93.24}\\ \textbf{$\pm$ 0.52}\end{tabular}  & 
\begin{tabular}[c]{@{}c@{}}\textbf{93.08}\\ \textbf{$\pm$ 0.50}\end{tabular}  & 
\begin{tabular}[c]{@{}c@{}}\textbf{93.15}\\ \textbf{$\pm$ 0.49}\end{tabular}  & 
\begin{tabular}[c]{@{}c@{}}\textbf{96.69}\\ \textbf{$\pm$ 0.35}\end{tabular}  &
\begin{tabular}[c]{@{}c@{}}\textbf{96.92}\\ \textbf{$\pm$ 0.38}\end{tabular}  &
\begin{tabular}[c]{@{}c@{}}\textbf{96.95}\\ \textbf{$\pm$ 0.34}\end{tabular}  &
\begin{tabular}[c]{@{}c@{}}\textbf{96.93}\\ \textbf{$\pm$ 0.35}\end{tabular} \\ \cline{2-9} &
\begin{tabular}[c]{@{}c@{}}\textbf{93.32}\\ \textbf{$\pm$ 0.68}\end{tabular}  & 
\begin{tabular}[c]{@{}c@{}}\textbf{93.80}\\ \textbf{$\pm$ 0.64}\end{tabular}  & 
\begin{tabular}[c]{@{}c@{}}\textbf{93.74}\\ \textbf{$\pm$ 0.66}\end{tabular}  & 
\begin{tabular}[c]{@{}c@{}}\textbf{93.76}\\ \textbf{$\pm$ 0.64}\end{tabular}  & 
\begin{tabular}[c]{@{}c@{}}\textbf{96.21}\\ \textbf{$\pm$ 0.49}\end{tabular}  &
\begin{tabular}[c]{@{}c@{}}\textbf{96.51}\\ \textbf{$\pm$ 0.45}\end{tabular}  &
\begin{tabular}[c]{@{}c@{}}\textbf{96.40}\\ \textbf{$\pm$ 0.44}\end{tabular}  &
\begin{tabular}[c]{@{}c@{}}\textbf{96.45}\\ \textbf{$\pm$ 0.44}\end{tabular} \\  \hline
\end{tabular}
}
\end{center}
\end{table}

\setlength{\tabcolsep}{4pt}
\begin{table}[t]
\centering
\begin{center}
\caption{Feature selection method comparison --- Indian Pines.}
\resizebox{\columnwidth}{!}{
\label{tab:rIP}
 \def\arraystretch{1.2}%
\begin{tabular}{|c|c|c|c|c|}
\hline
\multirow{2}{*}{\textbf{Method}} & \multicolumn{4}{c|}{\textbf{5   bands}} \\ \cline{2-5} 
 & \textbf{OA} & \textbf{Prec} & \textbf{Rec} & \textbf{F1} \\ \hline
\multirow{2}{*}{\textbf{PLS-DA}} & 96.68 $\pm$ 0.86 & 96.83 $\pm$ 0.99 & 95.62 $\pm$ 0.94 & 96.11 $\pm$ 0.74 \\ \cline{2-5} 
 & 97.17 $\pm$ 0.60 & 97.30 $\pm$ 0.79 & 96.66 $\pm$ 1.03 & 96.90 $\pm$ 0.84 \\ \hline
\multirow{2}{*}{\textbf{OCF}} & 96.68 $\pm$ 0.56 & 97.34 $\pm$ 0.76 & 96.34 $\pm$ 0.98 & 96.77 $\pm$ 0.82 \\ \cline{2-5} 
 & 97.02 $\pm$ 0.58 & 97.73 $\pm$ 0.51 & 97.14 $\pm$ 0.63 & 97.39 $\pm$ 0.48 \\ \hline
\multirow{2}{*}{\textbf{HAGRID}} & 96.74 $\pm$ 0.54 & 97.06 $\pm$ 0.75 & 96.34 $\pm$ 1.03 & 96.65 $\pm$ 0.88 \\ \cline{2-5} 
 & 97.03 $\pm$ 0.75 & 97.24 $\pm$ 0.86 & 96.72 $\pm$ 1.45 & 96.91 $\pm$ 1.21 \\ \hline
 
\multirow{2}{*}{\textbf{FNGBS}} & 97.49 $\pm$ 0.34 & 97.86 $\pm$ 0.36 & 97.64 $\pm$ 0.71 & 97.72 $\pm$ 0.5 \\ \cline{2-5} 
 & 97.34 $\pm$ 0.65 & 97.94 $\pm$ 0.52 & 97.75 $\pm$ 0.46 & 97.82 $\pm$ 0.44 \\ \hline
\multirow{2}{*}{\textbf{GSS}} & \textbf{98.08 $\pm$ 0.43} & \textbf{98.26 $\pm$ 0.42} & \textbf{98.39 $\pm$ 0.43} & \textbf{98.32 $\pm$ 0.39} \\ \cline{2-5} 
 & \textbf{98.24 $\pm$   0.39} & \textbf{98.56 $\pm$ 0.38} & \textbf{98.43 $\pm$ 0.42} & \textbf{98.48 $\pm$ 0.36} \\ \hline
\end{tabular} 
}
\end{center}
\end{table}

\section{Discussion} \label{discussion}

Using our IBRA method, we identified sets of influencing spectral bands for both datasets. These pre-selected bands explain the variance of their neighbors in the original spectrum with a VIF value greater than a threshold $\theta\in [5, 12]$; therefore, keeping them and removing the other bands allowed us to avoid spectral bands that did not contain useful information for performing classification. That is, our method effectively identifies those spectral bands that carry information for performing classification while discarding redundant spectral bands. Results shown in Table~\ref{tab:base} demonstrate that it is possible for a model trained on the subset of spectral bands determined by IBRA, to achieve high accuracy values ($\sim$ 97--99\%) similar to those obtained when using the full spectrum.

Our GSS method uses information entropy to identify which bands are more relevant among the pre-selected bands. However, if we need to select at most $k$ bands, the subset of bands with the greatest saliency values may not be the best selection. For instance, for the Kochia dataset, if $k=6$, the wavelengths of the bands with the highest entropy values are $[391.2, 463.3, 518.4, 616.0 , 658.4, 675.3]$; however, line \ref{line:vif} in Algorithm~\ref{alg:selection} detects strong multicollinearity between wavelengths $616.0, 658.4,$ and $675.3$. Rather than using redundant bands, we select a more diverse subset if this helps to improve the classification performance.      

From Tables \ref{tab:rkochia} and \ref{tab:rIP}, we see that our method achieved the highest performance on both datasets, which confirms our hypothesis. The results remain consistent even when considering different dataset sizes. In addition, Table~\ref{tab:rkochia} shows that there is a more noticeable gap in performance between our method and the others when selecting ten bands, rather than when selecting six bands. This confirms that the fewer spectral bands we select, the harder the task will be; however, multispectral imagers generally become more practical computationally as the number of spectral channels becomes smaller. 

Finally, it is worth noting that, with our method, the classification performance resulting from the ``original band selection" approach is very similar to the performance with the ``multispectral filter simulation" approach, unlike some of the compared methods. This can be explained by the way we selected the first band candidates using IBRA. That is, a spectral band corresponding to a local minimum in the plot of spectral index vs. distance $d(x)$ (Fig.\ref{fig:dist}) acts as a centroid because it is similar to the spectral bands located on either side. Therefore, if we take this local minimum as the central wavelength of a multispectral filter, generate a Gaussian distribution around it by considering a standard bandwidth, and integrate under the curve, then we obtain reflectance values similar to those of the central band. Given that the simulated multispectral filter and the original spectral band present similar information, their classification performance is similar. This is convenient for a multispectral sensor design, as we would like the central wavelength of a filter to be the most representative. 

\section{Conclusion} \label{conclusion}

Data captured by an imaging system is often processed to make observations and classifications about the world around us. The spatial and spectral content of the images obtained is key to analyzing the data, with spectral content playing a central role. However, the dense spectral information collected by a hyperspectral imager is not required for every application, and managing such data can be computationally expensive. The ability to determine the most relevant wavelengths for a given application enables using simpler multispectral imagers in place of hyperspectral imagers. This simplification would not only lead to economic savings, as fewer specialized storage and processing devices are required but also clarity and time-savings when analyzing data.

To allow for this simplification, we presented a method for selecting salient wavelengths from a hyperspectral data cube based on two main steps: A pre-selection step that identifies a subset of independent spectral bands (IBRA) and a final greedy selection step based on information entropy (GSS). Experimental results showed our band selection method generally outperformed other commonly-employed feature selection methods on the Kochia and Indian Pines datasets. 

Finally, we showed that the inter-band redundancy method does not only reduce the search space considerably, but it also provides potential filter centers that are suitable for the design of multispectral sensors. Hence, another outcome of this work is the aid in the design of compact multispectral imagers that will assist in applications such as automatically identifying herbicide-resistance biotypes of the weed kochia.

In the future, we plan to explore incorporating an attention mechanism into our CNN architecture. When combined with the band selection method proposed here, we expect further computational savings by enabling a more adaptive approach to using the selected bands. We also plan to explore implementation issues associated with developing multispectral sensors based on the designs recommended through these methods.

\balance
\bibliographystyle{IEEEtran}
\bibliography{references}

\end{document}